\documentclass[preprint,showpacs,showkeys,amsmath,amssymb,aps,doublespace]{revtex4}
\usepackage[dvips]{graphicx}
\usepackage{setspace}
\usepackage{amsmath}
\usepackage{amsfonts}
\usepackage{amssymb,color}

\begin{document}

\title{Pair contact process with diffusion of pairs}

\author{F. L. Santos$^{1}$, Ronald Dickman, and U. L. Fulco$^{1,3}$}

\affiliation{$^{1}$Departamento de Biof\'{\i}sica e Farmacologia, Universidade Federal do Rio Grande do
Norte, 59072-970, Natal-RN, Brazil}

\affiliation{$^2$Departamento de F\'{\i}sica, ICEx, and National Institute of Science and Technology
for Complex Systems,
Universidade Federal de Minas Gerais, 30123-970,
Belo Horizonte-MG, Brazil}

\affiliation{$^{3}$Departamento de F\'{\i}sica, Universidade Federal do Cear\'{a}, 60455-760,
Fortaleza-CE, Brazil}

\date{\today}

\begin{abstract}

The pair contact process (PCP) is a nonequilibrium stochastic model which,
like the basic contact process (CP), exhibits a phase transition to an absorbing state.
The two models belong to the directed percolation (DP) universality class, despite
the fact that the PCP possesses infinitely many absorbing configurations whereas
the CP has but one.  The critical behavior of the PCP with hopping by particles
(PCPD) is as yet unclear.
Here we study a version of the PCP in which nearest-neighbor particle {\it pairs} can hop
but individual particles cannot. Using
quasistationary simulations for three values of the diffusion probability ($D=0.1$, $0.5$ and
$0.9$), we find convincing evidence of
DP-like critical behavior.

\end{abstract}


\maketitle

\section{Introduction}

The exploration of universality classes associated with nonequilibrium phase transitions
continues to attract much interest \cite{marro,henkel,odor08}.
In the case of systems exhibiting a phase transitions to an absorbing state,
the generic universality class is that of directed percolation (DP)
\cite{grassberg,janssen}.
For example, both the basic contact process (CP) \cite{harris} and the
pair contact process (PCP) \cite{jensen,jensen1} belong to this class, despite the
fact that the former has a unique absorbing configuration while the latter possesses infinitely
many.
Allowing individual particles to hop in the PCP, one obtains the so-called pair
contact process with diffusion (PCPD). In this case there are only two absorbing states:
the empty lattice, and the state of a single particle hopping.

A model with the basic features of the PCPD was first proposed by Grassberger in 1982.  More
recent interest in this problem was stimulated by the work of Howard and
T\"auber \cite{howard} who discussed a generalized version, with a Langevin description
involving complex noise.  On the basis
of numerical results, Carlon
\textit{et al.} \cite{carlon}, suggested that the critical behavior of the PCPD would
fall in the parity-conserving
(PC) class.  Subsequent works \cite{hinrichsen,odor} suggested a different kind of critical behavior,
possibly masked by huge corrections to scaling.
(For a comprehensive review of analyses of the PCPD up to 2004, see the article
by Henkel and Hinrichsen \cite{hinrichsen3}.)

Given the controversies surrounding the PCPD, it seems worth checking whether allowing hopping
by pairs ({\it only}) changes the critical behavior of the PCP.  One expects
such a model to remain in the DP universality class, given that the
critical behavior of the CP is robust to the inclusion of nearest-neighbor
hopping \cite{diffCP}. Note as well that pairwise diffusion does not change the set of
absorbing configurations in the PCP.
In the present work, we study a variant of the PCP in which nearest neighbor pairs of particles
may hop together, while isolated particles cannot hop.
Using quasistationary (QS) simulations, we
determine several critical parameters for three values
of the diffusion probability: $D=0.1, 0.5$ and $0.9$. The balance of this paper is organized
as follows. In Sec. II we define the model and detail our simulation method. In
Sec. III we present our results; Sec. IV is devoted to discussion and conclusions.

\section{Model and Simulation Method}

The CP \cite{harris,marro} is one of the simplest and most studied models belonging to the DP
universality class. In the CP, each site $i$ of a lattice is either occupied
$[\sigma_i(t)=1]$ or vacant $[\sigma_i(t)=0]$. Transitions from $\sigma_i=1$
to $0$ occur at a rate of unity, independent of the neighboring sites. The
reverse transition can only occur if at least one neighbor is occupied: the
transition from $\sigma_i=0$ to $1$ occurs at rate $\lambda m$, where $m$
is the fraction of nearest neighbors of site $i$ that are occupied; thus the
state $\sigma_i=0$ for all $i$ is absorbing. $\lambda$ is the only control
parameter in the basic CP; the order parameter $\rho$ is the fraction of occupied
sites.

The PCP \cite{jensen,jensen1} is defined on a $d$-dimensional lattice, with each site again either
occupied or vacant. Only pairs of particles occupying nearest-neighbor sites exhibit
activity: each such pair has a rate $1-p$ of mutual annihilation, and a rate $p$
to create a new particle at a randomly chosen site neighboring the pair, if this
site is vacant. In the PCPD \cite{carlon},
in addition to the creation and annihilation processes present in the PCP,
each particle attempts to hop, at rate $D$, to a randomly chosen
nearest-neighbor (NN) site; the move is accepted if the target site is vacant.
Several variants
of the model, differing in how each process (creation, annihilation or diffusion) is
selected, have been studied \cite{hinrichsen3}.

The model studied here, pair contact process with diffusion of pairs (PCPDP), again features the
pair-mediated annihilation and creation processes of the PCP.  In addition, a fraction $D$
of all events are hopping attempts by a NN particle pair.  The model is defined on a ring
of $L$ sites. The transition rates associated with each pair $(AA$) are:

\begin{equation}
\begin{array}{cc}
\emptyset AA \rightarrow AA \emptyset, \;\;\;
AA \emptyset \rightarrow \emptyset  AA, & \mbox{at rate $D/2$} \\
\\
AA \emptyset, \;\; \emptyset AA \rightarrow AAA, & \mbox{at rate $p(1-D)/2$} \\
\\
AA \rightarrow \emptyset \emptyset, & \mbox{at rate $(1-p)(1-D)$.}
\end{array}
\end{equation}

\noindent Thus hopping ceases, along with all other activity, in the absence of pairs.

The most obvious definition of the order parameter in the PCPDP is the pair density $\rho_p$,
that is, the number of nearest-neighbor occupied pairs divided by the total number
of nearest-neighbor pairs on the lattice (on the ring, the latter is equal to the number of sites).
Since the number of particles can only change in the presence of pairs,
one might expect the {\it excess} particle density, $\Delta \rho_{part} = \rho_{part} - \rho_{part,c}$,
to scale in a similar manner.  (Note that the critical particle density $\rho_{part,c}$
is nonzero \cite{jensen}.)
We confirm that $\Delta \rho_{part}$ and $\rho_p$ exhibit similar scaling properties.

We sample the quasistationary (QS) distribution of the
process, i.e., conditioned on survival, using a simulation
method that yields quasistationary properties directly \cite{dickman3}. This is done by
maintaining, and gradually updating, a set of configurations visited during the evolution;
when a transition to the absorbing state is imminent the systems is instead placed in
one of the saved configurations. Otherwise the evolution is identical to that of a conventional
simulation. The set of saved configurations is updated by replacing with a small probability, $p_{rep}$,
at each time step, one of the saved configurations with the current one.

We perform extensive simulations of the one-dimensional PCPDP, using systems of
$L=10, 20, 40, 80, 160, 320, 640$ sites, using the QS simulation method. Each realization
of the process is initialized with all sites occupied, and runs for $5\times10^6$ to
$10^8$ time steps (longer runs for larger systems).
Our results are calculated over samples sizes $10$-$20$ realizations.
The number of saved configurations ranges from 100 to 1000 (larger numbers for smaller systems).
Values of $p_{rep}$ range from
$10^{-3}$ to $5\times10^{-4}$ (smaller values for larger systems).
During the relaxation phase we use a $p_{rep}$
ten times larger, to eliminate the influence of the initial configuration.
Following this relaxation phase, we accumulate histograms of the time during
which the system has exactly $1,2,...,n,...$ pairs, and similarly for particles.
Using the histograms we calculate the densities of pairs and of particles, the
moment ratio $m \equiv \langle \rho_{p}^2 \rangle/\langle \rho_{p} \rangle ^2$
and the reduced fourth cumulant
$q_4 \equiv K_4/K_2^2$, where
$K_2=  \mbox{var}(\rho_p)$ and,

\begin{equation}
K_4= \langle \rho_p^4 \rangle  -4\langle \rho_p^3 \rangle \langle \rho_p \rangle
-3 \langle \rho_p^2 \rangle^2 + 12 \langle \rho_p^2 \rangle \langle \rho_p \rangle^2
-6 \langle \rho_p \rangle^4.
\end{equation}

\noindent The QS lifetime $\tau$ is taken as the mean time
between attempts to visit an absorbing configuration.

\section{Simulation Results}

\begin{figure}[!hbt]
\includegraphics[clip,angle=0,width=0.8\hsize]{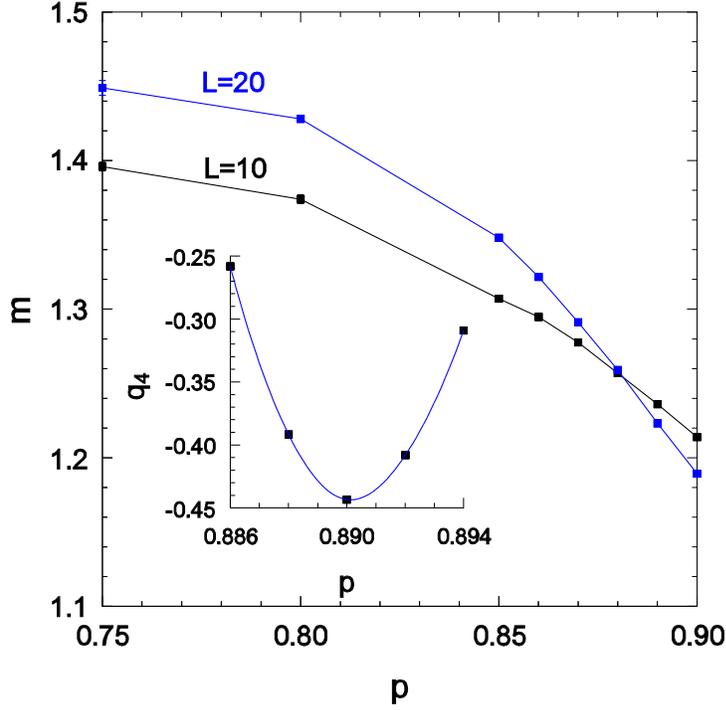}
\caption{(Color online) Moment ratio $m$ versus creation parameter
$p$ for $D=0.5$, $L=10$, 20.  Inset: Reduced fourth cumulant versus $p$ for $L=160$, $D=0.5$.}
\label{fig1}
\end{figure}

For each diffusion rate studied,
the first step in the analysis is to determine the critical annihilation probability
$p_c(D)$. Experience with absorbing-state phase transitions leads us
to expect the following scaling properties at the critical
point: $\rho_p \sim L^{-\beta/\nu_\perp}$; $\tau\sim L^{z}$;
and $m(L) \to m_c$, a universal critical value \cite{dickman1}.
We obtain a preliminary estimate of $p_c$
from the crossing of the moment ratios $m$ for system sizes $L=10$ and 20,
and then concentrate our efforts in simulations near this value.
(For $D=0.5$, for example, we
focus on the interval $p \in [0.85,0.90]$,
see Fig. \ref{fig1}).

\begin{figure}[!hbt]
\includegraphics[clip,angle=0,width=0.8\hsize]{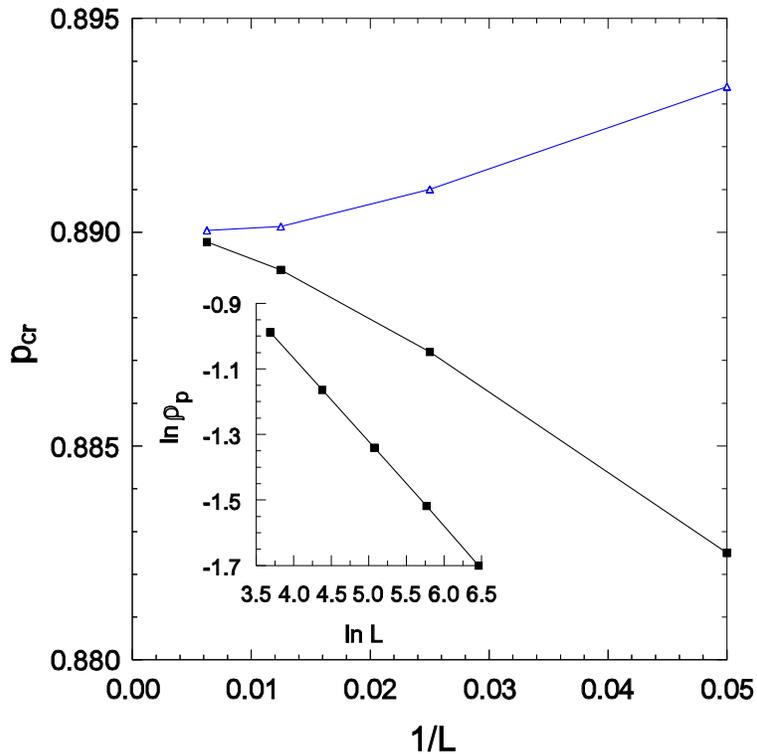}
\caption{(Color online) Values $p_\times$ and $p_q$ associated, respectively, with moment ratio crossings
(lower) and minima in the reduced fourth cumulant (upper), versus $1/L$, for $D=0.5$.
Inset: Order parameter $\rho_p$ versus system size, for $D=0.5$ and $p=0.89$.}
\label{fig2}
\end{figure}

For each pair of consecutive lattice sizes, $L$ and
$2L$, we determine the crossing values $p_\times$ and $m_\times$.
An independent estimate of the critical point is afforded by the reduced fourth cumulant, $q_4$,
which takes a pronounced minimum at a value $p_q(L)$ that converges to the critical
value; see Fig. \ref{fig1}, inset. In Fig. \ref{fig2} we plot
$p_\times$ and $p_q$
versus $1/L$, for $D=0.5$; both sets of
values converge quickly to a limit which we estimate, using quadratic extrapolation
(versus $1/L^2$), as $p_c=0.8900(2)$.

The finite-size scaling relations for $\rho_p$, $\tau$, and $m$, cited above, are satisfied
to good precision for $p = 0.8900$; the data for the order parameter are shown in the inset of
Fig. \ref{fig2}.
Including results for $L=640$, however, we find small but significant curvatures in the
graphs of $\ln \rho_p$ and $\ln \tau$ versus $\ln L$, indicating that $p_c$ is slightly
larger than 0.8900.  (The curvatures are: -0.0019(4) in the case of $\rho_p$
and -0.009(3) for $\tau$.)
We perform additional simulations for $L=640$ at $p=0.8903$ and 0.8906, and then calculate
the curvatures of the graphs of $\ln \rho_p$ and $\ln \tau$ versus $\ln L$ for diverse
values of $p$ in the interval [0.89, 0.8903], using polynomial interpolation of the simulation
data as necessary, to estimate the quantities of interest at intermediate points.  We observe
no significant curvature for $p$ in the interval [0.89004, 0.89010], leading to
our final estimate, $p_c = 0.89007(3)$, for $D=0.5$.  The associated critical parameters are
$\beta/\nu_\perp = 0.252(2)$, $z=1.573(10)$, and $m=1.1758(24)$.  (Note that the principal
source of uncertainty in these estimates is the uncertainty in $p_c$ itself.)

The data for the particle density $\rho_{part}$ scales in the same manner as the pair density:
at the critical point, we find, to good precision, $\rho_{part} \simeq \rho_{part,c} + A L^{-\beta/\nu_\perp}$,
using the same value for $\beta/\nu_\perp$ as found in the analysis of the pair density.
(For $D=0.5$, for example, the limiting particle density is $\rho_{part,c} = 0.2321(2)$.)
The moment ratio associated with the particle density,
$m_{part} \equiv \langle (\rho_{part}-\rho_{part,c})^2 \rangle/\langle \rho_{part}-\rho_{part,c} \rangle^2$,
behaves similarly to the moment ratio associated with pairs; for $D=0.5$ we find $m_{part,c} = 1.177(5)$.

\begin{figure}[!hbt]
\includegraphics[clip,angle=0,width=0.8\hsize]{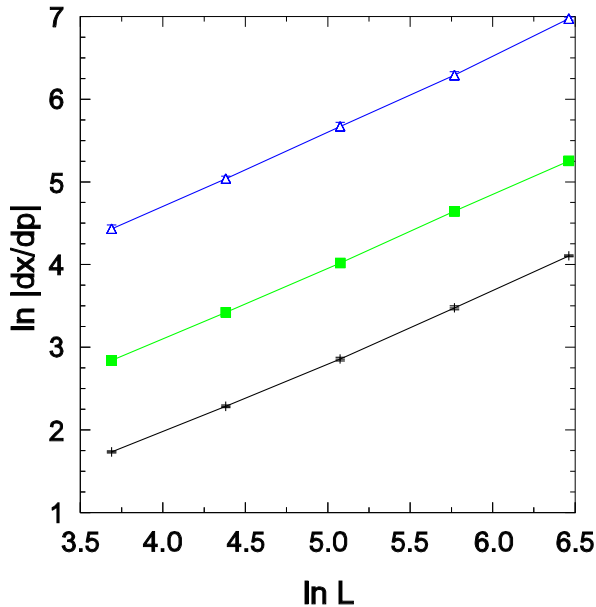}
\caption{(Color online) Derivatives $|dm/dp|$ (lower), $d \ln \tau/dp$ (upper), and
$d \ln \rho_{p}/dp$ (middle) versus $L$ for $D=0.5$.}
\label{derivs}
\end{figure}

The simulation data also permit a direct estimate of the exponent $\nu_\perp$, albeit
with somewhat limited precision.
Finite-size scaling implies that the derivatives
$|dm/dp|$, $d \ln \tau/dp$ and $d \ln \rho_{p}/dp$, evaluated at the critical point,
follow $| dx /dp| \propto L^{1/\nu_\perp}$ (here $x$ stands for any of the quantities mentioned).
We estimate the derivatives via least-squares linear fits to the data on an interval
that includes $p_c$. (The intervals are small enough that the graphs show no significant
curvature.)  Power-law dependence of the derivatives on system size is verified in
Fig. \ref{derivs}.  Linear fits to the data for the three largest sizes, for $m$, $\ln \rho_p$,
and $\ln \tau$ yield
$1/\nu_\perp = 0.900(4)$, 0.940(26) e 0.891(7), respectively, leading to the estimate
$\nu_\perp = 1.10(2)$.  The simulations and analyses described above were repeated for diffusion
rates $D=0.1$ and 0.9, yielding the results shown in Table I.  For $D=0.9$,
finite-size corrections are much stronger than for the smaller diffusion rates, and we
found it necessary to extend the study to larger system sizes to obtain reliable results.
The values listed in Table I are based on studies using $L=320$, 640, 1280, and 2560,
in this case.

\begin{table}[ht]
\begin{center}
\begin{tabular}{cccccc}
\hline \hline
D     &~~~ $p_c$          &~~~ $\beta/\nu$ &~~~ $z$         &~~~ $\nu_\perp$ &~~~ $m_p$ \\
\hline
$0.1$ &~~~ $0.91720(1)$   &~~~ $0.252(1)$  &~~~ $1.584(11)$ &~~~ $1.11(4)$   &~~~  $1.173(1)$ \\
$0.5$ &~~~ $0.89007(3)$   &~~~ $0.252(2)$  &~~~ $1.573(10)$ &~~~ $1.10(2)$   &~~~  $1.1758(2)$ \\
$0.9$ &~~~ $0.83470(5)$   &~~~ $0.253(1)$  &~~~ $1.573(5)$  &~~~ $1.12(3)$   &~~~  $1.178(2)$ \\
\hline
DP    &~~~                &~~~  $0.25208(5)$ &~~~ $1.5807(1)$ &~~~ $1.096854(4)$ &~~~ $1.1736(1)$ \\
\hline\hline
\end{tabular}
\end{center}
\caption{Critical exponent values for the PCPDP and DP; figures in
parentheses denote uncertainties. DP values from Refs. \cite{dickman1} and \cite{jensen2}.}
\end{table}

\section{Conclusions}

We study a version
of the pair contact process in which nearest-neighbor particle
pairs (but not isolated particles) diffuse.
The results of quasistationary simulations, for $\beta/\nu_\perp$, $z$, $\nu_\perp$,
and the moment ratio $m$, confirm to good precision
that the scaling properties of this model coincide with those of
directed percolation.
This conclusion is what one would expect on the basis of universality:
diffusion of pairs changes nothing fundamental in the propagation of activity
or in the space of absorbing configurations of the original pair contact
process.  The situation is quite different from that of the PCP with
diffusion of {\it particles}: the critical behavior of the PCPD is still not
fully understood, and is evidently subject to much stronger corrections to
scaling than is the model studied here (PCPDP).  Thus our conclusion
that the PCPDP belongs to the directed percolation universality class does
not imply that the same holds (or does not hold) for the PCPD.

\begin{acknowledgments}
This work was partially financed by the Brazilian Research
Agencies CAPES (PROCAD and Rede NanoBioTec), CNPq (Edital Universal)
and FAPERN/CNPq (Pronex).
\end{acknowledgments}

\end{document}